\documentclass[conference]{IEEEtran}
\IEEEoverridecommandlockouts

\usepackage{enumitem}    

\usepackage[table,dvipsnames]{xcolor}
\usepackage{enumitem}    
\newif\ifoutline
\outlinetrue

\usepackage{multirow}
\usepackage{multicol}
\usepackage{booktabs}
\usepackage{gensymb}

\usepackage{cite}
\usepackage{amsmath,amssymb,amsfonts}
\usepackage{algorithmic}
\usepackage{graphicx}
\usepackage{textcomp}
\usepackage{xcolor}
\usepackage{mathtools}
\usepackage{subcaption}

\usepackage{listings}
\usepackage{xcolor}
\usepackage{makecell}
\usepackage{multirow}
\usepackage{pifont}
\usepackage{threeparttable}  

\definecolor{codegreen}{rgb}{0,0.6,0}
\definecolor{codegray}{rgb}{0.5,0.5,0.5}
\definecolor{codepurple}{rgb}{0.58,0,0.82}
\definecolor{backcolour}{rgb}{0.95,0.95,0.92}

\lstdefinestyle{mystyle}{
    backgroundcolor=\color{backcolour},   
    commentstyle=\color{codegreen},
    keywordstyle=\color{magenta},
    numberstyle=\tiny\color{codegray},
    stringstyle=\color{codepurple},
    basicstyle=\ttfamily\footnotesize,
    breakatwhitespace=false,         
    breaklines=true,                 
    captionpos=b,                    
    keepspaces=true,                 
    numbers=left,                    
    numbersep=5pt,                  
    showspaces=false,                
    showstringspaces=false,
    showtabs=false,                  
    tabsize=2
}

\usepackage[thinc]{esdiff}

\usepackage{tikz}
\newcommand*\circled[1]{\tikz[baseline=(char.base)]{ \node[shape=circle,fill=black,inner sep=0.4pt] (char) {\textcolor{white}{#1}};}}

\def\BibTeX{{\rm B\kern-.05em{\sc i\kern-.025em b}\kern-.08em
    T\kern-.1667em\lower.7ex\hbox{E}\kern-.125emX}}

\newcommand{\cmark}{\textcolor{OliveGreen}{\ding{51}}} 
\newcommand{\xmark}{\textcolor{BrickRed}{\ding{55}}}   

\begin{document}

\title{


A Unified Framework for Mapping and Synthesis of Approximate R-Blocks CGRAs

}

\author{
    \IEEEauthorblockN{
        Georgios Alexandris\textsuperscript{1}, 
        Panagiotis Chaidos\textsuperscript{1}, 
        Alexis Maras\textsuperscript{1}, 
        Barry de Bruin\textsuperscript{2}, 
        Manil Dev Gomony\textsuperscript{2}, 
        Henk Corporaal\textsuperscript{2},\\
        Dimitrios Soudris\textsuperscript{1},
        Sotirios Xydis\textsuperscript{1}
    }
    \IEEEauthorblockA{
        \textsuperscript{1}National Technical University of Athens, Greece\\
        \textsuperscript{2}Eindhoven University of Technology, The Netherlands \\\{galexandris,pchaidos,amaras,dsoudris,sxydis\}@microlab.ntua.gr, \{e.d.bruin,m.gomony,h.corporaal\}@tue.nl
    }
 \thanks{This work is funded in part by the Convolve project evaluated by the
    EU Horizon Europe research and innovation program under grant agreement
    No. 101070374}
}

\maketitle

\begin{abstract}

The ever-increasing complexity and operational diversity of modern Neural Networks (NNs) have caused the need for low-power and, at the same time, high-performance edge devices for AI applications. Coarse Grained Reconfigurable Architectures (CGRAs) form a promising design paradigm to address these challenges, delivering a close-to-ASIC performance while allowing for hardware programmability. In this paper, we introduce a novel end-to-end exploration and synthesis framework for approximate CGRA processors that enables transparent and optimized integration and mapping of state-of-the-art approximate multiplication components into CGRAs. Our methodology introduces a per-channel exploration strategy that maps specific output features onto approximate components based on accuracy degradation constraints. This enables the optimization of the system's energy consumption while retaining the accuracy above a certain threshold. At the circuit level, the integration of approximate components enables the creation of voltage islands that operate at reduced voltage levels, which is attributed to their inherently shorter critical paths. This key enabler allows us to effectively reduce the overall power consumption by an average of 30\% across our analyzed architectures, compared to their baseline counterparts, while incurring only a minimal 2\% area overhead. The proposed methodology was evaluated on a widely used NN model, MobileNetV2, on the ImageNet dataset, demonstrating that the generated architectures can deliver up to 440 GOPS/W with relatively small output error during inference, outperforming several State-of-the-Art CGRA architectures in terms of throughput and energy efficiency.
\end{abstract}


\section{Introduction}

The rapid growth of Artificial Intelligence (AI) in modern computing has led to several attempts at deploying such applications in Edge computing environments. Edge AI paradigms, however, increasingly depend on both fine-tuning Neural Network (NN) models and the development of specialized hardware architectures that enforce advanced levels of energy efficiency \cite{aisurvey}. 

\par Despite the resource constraints of edge devices, research has placed a lot of effort into producing high-performance accelerators capable of achieving very promising results. However, the architectural diversity of modern Deep Neural Networks (DNNs) highlights the need for reprogrammable hardware so that accelerators can adapt to dynamic changes and techniques such as layer fusion, retraining, and other model-level modifications. To ensure compatibility with this dynamically changing environment, use-case specific accelerators exist, targeting Neural Network implementations \cite{edgetpu}, \cite{gemmini} that tend to be more programmable and configurable, with the added cost of performance loss.

\begin{table*}[t]
    \centering
    \renewcommand{\arraystretch}{1.2} 
    \setlength{\tabcolsep}{6pt}
    \small
    \begin{tabular}{c|ccccc}
         CGRA Architecture &  Approximate Computing & Voltage Scaling &  Flexible PEs &  \makecell{High Level \\ Programmability} & DNN Support \\ \hline
         X-CGRA\cite{akbari_x-cgra_2020}&  \cmark & Static & \xmark &  DFG-Based&\xmark\\
         GREEN\cite{green_2024}&  \cmark & \xmark & \xmark &  DFG-Based&\xmark\\
         Brandalero et al.\cite{approxonthefly}& \cmark & \xmark & \cmark &  Behavioural Model&\xmark\\
         CGRA-ME\cite{cgrame}&   \xmark & \xmark & \xmark &  DFG-Based&\xmark\\
         REVAMP\cite{revamp}&   \xmark & \xmark &  \cmark &  DFG-Based&\xmark\\
         CGRA4ML\cite{cgra4ml}&   \xmark & \xmark & \xmark &  Custom Python Flow&\cmark\\
         ML-CGRA\cite{mlcgra} & \xmark & \xmark & \xmark & MLIR from Python&\cmark\\
         ICED\cite{iced}& \xmark & Dynamic & \xmark & DFG-Based&\xmark\\
         R-Blocks\cite{de_bruin_r-blocks_2024}& \xmark & \xmark & \cmark & C compiler&\xmark\\
         \textbf{Our Work}&  \cmark &Static&  \cmark &  C compiler&\cmark\\
    \end{tabular}
    \caption{Qualitative comparison among the state-of-the-art CGRAs}
    \label{tab:comparison}
\end{table*}

\par Coarse-Grained Reconfigurable Processor Architectures (CGRAs) provide a great alternative to ASIC implementations, utilizing application-specific functional units with a reconfigurable interface between them \cite{7818353}. CGRA-ME\cite{cgrame} forms such a typical CGRA architectural scheme, with high-level configuration and DFG programmability of a 2D Processing Element grid. To maximize energy efficiency, more aggressive design methods, e.g., approximate computing, have been adopted to further improve latency and energy for such devices. Research works such as X-CGRA\cite{akbari_x-cgra_2020} and GREEN\cite{green_2024} introduce precision-scaled approximation, with X-CGRA allowing mapping of RISC-like instructions on the architecture and GREEN with extra support for MIMD-like code execution. Additionally, the work presented in Brandalero et al. \cite{approxonthefly} explores and generates different systolic-like Processing Elements (PEs), in which tiles capable of variable approximation (variance in aggressiveness) are connected to a shared memory, allowing for each one to be selected independently in the mapping phase. Recently, heterogeneous CGRAs have also been proposed \cite{revamp, de_bruin_r-blocks_2024}. The works of RAVAMP\cite{revamp} and R-Blocks\cite{de_bruin_r-blocks_2024} paved the way for demonstrating how disaggregated PEs should perform in a standardized formation, presented in RAVAMP or in fully reconfigurable one, which is handled in the work of the R-Blocks, providing also a high-level C compiler. Finally, ICED\cite{iced} introduces DVFS principles on standardized CGRA architectures\cite{opencgra}, with voltage scale-aware mapping of certain applications on the CGRA's PEs.

\par In this paper, we present a CGRA end-to-end synthesis framework that co-optimizes application mapping and CGRA hardware to achieve energy-efficient designs for edge AI applications. Our methodology leverages approximate computations to exploit the inherent error resiliency of DNNs. At first, voltage scaling is introduced on selected CGRA regions, e.g., the approximate components or the ALUs, etc., which exhibit higher timing slack compared to other CGRA counterparts. This selective voltage islands formation minimizes the overall energy consumption without inducing timing violations, ensuring a balance between energy efficiency and computational reliability. Furthermore, a seamless end-to-end flow is provided, starting from a high-level DNN model exploration compatible with the PyTorch library. More specifically, an extensive analysis of the impact on the accuracy of the model is introduced when applying approximate operations to specific output features and accordingly maps them to either the approximate or the accurate multiplication units, based on user-defined constraints such as the maximum allowed accuracy drop. The key contributions of our work are:

\begin{itemize}
    \item We extend R-Blocks micro-architecture \cite{de_bruin_r-blocks_2024} with state-of-the-art approximate computing capabilities. We integrate state-of-art approximate functional unit implementations, i.e. DRUM multiplier\cite{hashemi_drum_2015}, enabling more efficient execution of deep learning workloads.
    \item We further optimize the new CGRA architecture with a voltage island-based power optimization strategy that enables groups of CGRA's processing elements (PEs) to operate at different supply voltages. By strategically assigning voltage levels to PEs executing approximate and accurate operations, we enable effective cooperative optimization between approximate computing and voltage scaling, thereby aggressively minimizing energy consumption while maintaining computational integrity.
    \item We implement and analyze a novel end-to-end C-to-RTL mapping framework for approximate CGRAs, which co-explores resource allocation and QoS applied on modern DNN models. By systematically analyzing the importance of each DNN layer’s output features, our CGRA dynamically assigns computations to either accurate or approximate functional units, ensuring an optimal trade-off between computational accuracy and energy efficiency under user-defined constraints.

\end{itemize}

\par Through an extensive experimental evaluation conducted on a widely adopted Convolutional Neural Network (CNN) MobileNetV2, using the ImageNet dataset, we demonstrate the effectiveness of our proposed architecture in achieving substantial energy efficiency improvements. Our results indicate that our methodology achieves, on average, a 30\% reduction in energy consumption compared to the baseline R-Blocks architecture, all while maintaining relatively small error and without increasing computational resource requirements.
Furthermore, the proposed CGRA attains an energy efficiency in the range of 378 to 440 GOPS/W, significantly outperforming several state-of-the-art CGRA architectures and frameworks. These results highlight the effectiveness of our heterogeneous computing approach, which leverages approximate computing and voltage scaling techniques to enhance energy efficiency without compromising model accuracy.

\section{Related Work}
\label{sec:related}

\par Several research works have proposed energy-efficient CGRA accelerators. CGRA-ME \cite{cgrame} was among the first to introduce the concept of higher-level reconfigurability, making architectural exploration accessible to the user while providing a Data-Flow-Graph (DFG) to CGRA mapping framework. Based on this approach,  recent efforts, like X-CGRA\cite{akbari_x-cgra_2020}, explore CGRA architectures with flexible precision scaling on their Processing Elements (PEs). Their design flow leverages QoS constraints in order to map each DFG node from image processing benchmarks to minimize energy consumption during execution. 
Another noteworthy approach comes from the GREEN CGRA\cite{green_2024}, which provides support for both SIMD and MIMD instructions along with precision scaling and approximation arithmetic components. From an architectural standpoint, they employ uniform PE architectures (i.e., each PE in their array has the same functionality) and standardized square grid layouts across the CGRA fabric. 
The work presented by Brandalero et al.\cite{approxonthefly} implements tiles with varying approximation scales (different output quality) within the same row, sharing a common memory. Both \cite{green_2024} and \cite{approxonthefly} mapping strategy is based on costly iterative evaluation of trade-offs between energy savings and output degradation, to determine the final approximation units of the application execution.  

\par Several state-of-the-art works strive to decouple the memory and logic blocks from the CGRA. REVAMP CGRA\cite{revamp} builds upon the methodology of the HyCUBE CGRA\cite{hycube}, which generates grids of homogeneous PEs and routes the application DFG among them. They both utilize dedicated routing and placement principles to find an optimal mapping for each use case. Unlike HyCUBE, REVAMP introduces custom memory and logic tiles to improve energy and area and assigns separate PEs for multiplication, addition, and memory accesses. 

\par Voltage scaling in CGRAs was only recently proposed by the ICED\cite{iced} project. ICED applies voltage scaling by forming power islands, using different regions according to energy minimization and mapping constraints, compared with other state-of-the-art works such as HyCUBE CGRA \cite{hycube}. It also implemented a fully fledged flow,  starting from a compilation flow, generating standardized CGRAs\cite{opencgra}, propagating the generated DFG, and applying specific voltage island constraints, limiting the voltage scale accordingly to different regions of the produced ASIC.

\par Specifically for the Machine Learning (ML) domain, the CGRA4ML \cite{cgra4ml} project provides a high-level exploration framework based on the KERAS library, generating optimal mappings of DNN kernels on CGRA platforms alongside appropriate design constraints for both FPGA and ASIC synthesis flow. 
ML-CGRA\cite{mlcgra}, focus on optimizations on the compilation flow. By leveraging the MLIR\cite{mlir} dialect at a higher abstraction level, they are able to produce optimal DFG mappings targeting CGRA PE grids. Their work uses OpenCGRA\cite{opencgra} as its target platform, which features a more evenly distributed and homogeneous space to allocate all the necessary resources for each target ML model.

\par Summarizing the above discussion, Table \ref{tab:comparison} presents a qualitative comparison among key studies on CGRA implementations. We differentiate from prior art in several directions. First of all, we introduce a novel heterogeneous CGRA architecture specifically designed for DNN inference acceleration, setting itself apart from prior works through its co-optimization of approximate arithmetic, power management, and application mapping to achieve a fine-grained balance between performance, energy efficiency, and accuracy. Furthermore, we feature a high-level accuracy analysis framework that assesses the impact of approximate and precise multipliers on DNN inference quality. It enables a systematic application mapping strategy guided by QoS constraints, optimizing accuracy-power trade-offs, thus providing finer-grained control compared to prior CGRA-based ML accelerators. 

\section{SW Compilation \& HW Generation Flow}
\label{sec:axr-blocks}
\subsection{Architecture Overview}

\par
\begin{figure}[t]
  \centering
  \includegraphics[width=0.3\textwidth]{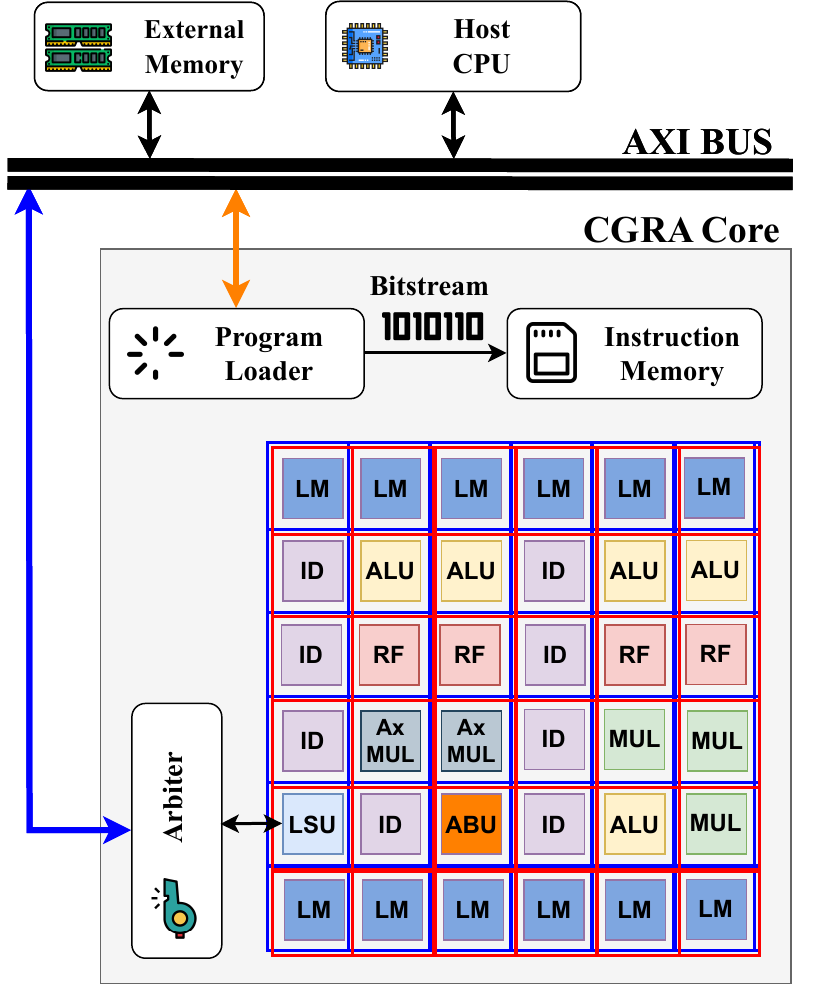}
  \caption{Overview of the architecture template and its integration with the host system.}
  \label{fig:arch}
\end{figure}

 Figure \ref{fig:arch} shows an abstract overview of the architectural organization of the proposed CGRA. We introduce a more heterogeneous approach, where each tile can be configured to perform different operations, while it also includes internal local SRAM memories to enable faster data transfers, i.e., enabling a form of near SRAM-memory computing. Furthermore, except the default operational PEs, there are also some approximation multiplication tiles introduced, mentioned as Ax MUL.
Each tile is connected to an Instruction Decode (ID) tile in the scalar version of the CGRA (SISD configuration), or a group of tiles can be controlled simultaneously by the same ID performing a vectorized operation, with operational vector size equal to the PEs contributing (SIMD configuration). All tiles are interconnected with two programmable Network-on-Chip (NoC) infrastructures: one for control words and one for data, transferring either instruction words or data segments according to the original architecture description (tile-to-tile direct communication). The overall programmable NoC infrastructure is organized in a standard 2D-mesh consisting of Wilton switchboxes\cite{wilton} for robust and scalable interconnection. The communication with the rest of the system is being handled through an AXI bus, which uses a Master Interface to access the external memory address spaces and a Slave Bus to load the executable program (i.e., CGRA bitstream) into the local instruction memory and configure the functional units. 

\subsection{Architecture and Code Mapping Analysis}
\label{subsec:AxR_mapping_flow}

\par
\begin{figure}[t]
  \centering
  \includegraphics[scale = 0.55]{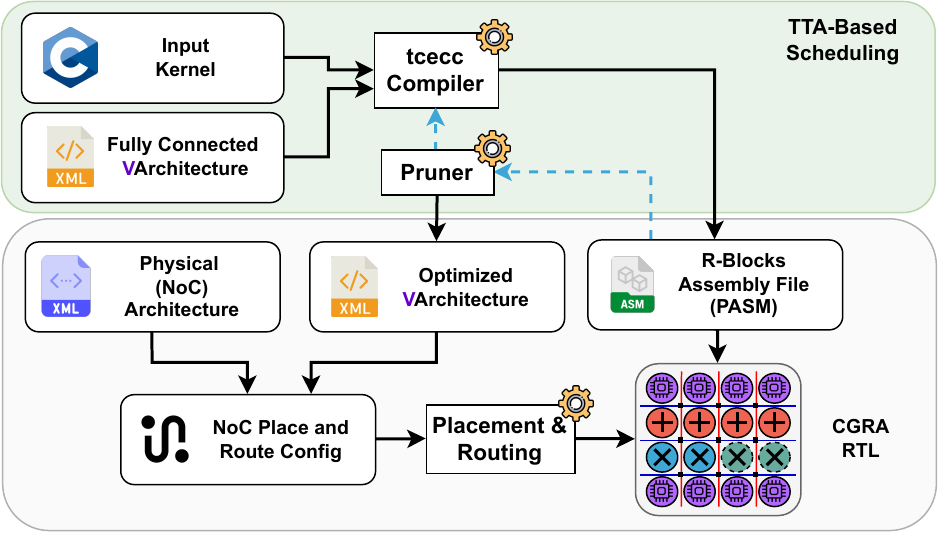}
  \caption{SW Compilation \& HW Generation Flow}
  \label{fig:mapping}
\end{figure}

 Figure \ref{fig:mapping} describes the compilation flow from the high-level application down to the final synthesizable RTL. The applications that can be scheduled onto our CGRA architecture are compiled from a high-level C code. The compilation process is based on OpenASIP's \cite{jaaskelainen2017hw} compiler (tcecc). This framework allows the CGRA to be modeled as a Transport-Triggered Architecture (TTA) machine \cite{tta}, where application scheduling is translated as a series of data and control transfers across time between all functional units. The output of this compilation flow is referred to as "Parallel Assembly" (PASM), which consists of a separate yet synchronized assembly instruction set for each functional unit, fetching one instruction per clock cycle for every PE. PASM gets translated into a binary file that directly programs/configures the CGRA's Instruction Memory(IM). During the execution phase, the IM is then accessed and decoded by each ID tile.

\par As shown in Figure \ref{fig:mapping}, the tcecc compiler takes into consideration a high-level (virtual) fully-connected micro-architectural model of the target CGRA, thus, it performs operation scheduling under relaxed inter-tile connectivity constraints. This high-level fully-connected micro-architectural model is then iteratively refined to fit the connectivity constraints imposed by the 2D-mesh programmable NoC. An iterative pruning process is employed, i.e., the Pruner (Figure \ref{fig:mapping}), which reroutes the control and the data transfers and then removes underutilized or redundant connections while maintaining the application's schedulability. 

\par Finally, the interconnection optimized virtual architecture together with the scheduled PASM are placed and routed onto the 2D-mesh grid of programmable NoC. 
This process maps each FU of the optimized virtual architecture onto the 2D CGRA grid, and then finds the optimal routing paths among the switchboxes for both the instruction and the data communication between tiles. The configuration of the NoC produced by this step, along with the executable program, allows us to produce our final synthesizable design and simulate it accordingly at the RTL and post-synthesis netlist level for validation. 


\subsection{Analysing and Integrating DRUM approximate multipliers}

\par Our framework embraces approximate computing by incorporating approximate arithmetic units in the CGRA computing fabric and then by offering selective operation mapping  (Section \ref{sec:approx_mapping} onto them in a disciplined manner to minimize energy consumption. We focus on approximate multiplication operators, since modern DNNs are computationally dominated by millions of Multiply and Accumulate operations. Without loss of generality, we utilize the DRUM approximate multiplication unit \cite{hashemi_drum_2015}, as our driver for approximate arithmetic. Several studies in the domain of approximate multiplier design \cite{microAX18, dacAX19, Jiang2020, TECS24Axsurvey} have shown the consistent effectiveness of DRUM multipliers. 
DRUM offers a competitive trade-off between area and power efficiency while maintaining acceptable accuracy loss, making it well-suited for deep learning applications tolerant to approximation errors. 

\par In brief, a DRUM$k$ multiplier takes two n-bit numbers as inputs. The rank of the Leading One bit from each operator gets captured, and then the other $k-1$ bits following ($k$ is a user-defined configuration parameter) get multiplied with an accurate $k$ $\times$ $k$ bit multiplier. Before the multiplication, the LSB from each $k$-bit captured operator becomes 1, for unbiasing. After the multiplication is completed, a barrel shifter corrects the position for the final output to have the correct power-of-2 correspondence. The remaining LSBs following this output are truncated to produce the final result.

\par Table \ref{tab:drum} presents an evaluation of the DRUM$k$ multiplier power, area, and delay, where the configuration parameter $k$ specifies the number of bits retained after the leading one in each operand. The DRUM multiplication units evaluated were the ones extended to ensure compatibility with the system's interface. The Root Mean Squared Error (RMSE) was measured by executing all possible combinations of 8$\times$8-bit multiplications, since we will mainly target use cases of INT8-quantized DNNs. Power and area metrics were obtained using Synopsys Design Compiler, synthesized at a 22nm technology node and a supply voltage of 0.8V. As expected, RMSE tends to decrease as the value of $k$ increases since a less aggressive approximation results in lower error rates. This drop in computational accuracy is capable of yielding trade-offs in all other domains of interest, namely power, area, and delay. Interestingly, Table \ref{tab:drum} shows that these gains demonstrate non-linearity. Very aggressive options, such as DRUM4, offer only marginal gains when compared with the rest of the DRUM configurations while incurring a disproportionately large increase in the error.

\begin{table}[h]
\centering
\small
\begin{tabular}{c|cccc}
MUL  & RMSE  &   Power ($\mu$W) & Area ($\mu$m$^2$) & Delay (ps) \\ \hline
DRUM4    & 385.4  & 294  & 430  & 797    \\
DRUM5    & 198.1  & 302  & 451  & 820     \\
DRUM6    & 101.3  & 315  & 475  & 883      \\
DRUM7    & 13.1   & 338  & 493  & 932      \\          
Accurate & 0      & 638  & 991  & 1540
\end{tabular}
\caption{DRUM Power-Performance-Area Analysis}
\label{tab:drum}
\end{table}

\vspace{-2ex}

\noindent  \textbf{HW/SW integration of DRUM:} DRUM is included in the CGRA as a standardized multiplication tile, supporting the same ISA micro-instructions and configuration principals for I/O format and operation delays. With this configuration, the mapping approach (Section \ref{subsec:AxR_mapping_flow}) can be utilized as is without significant changes. 

\begin{lstlisting}[language=C, numbers=none, caption=Example of intrinsic Driven Operations, label=list:intrinsic ]
if(condition){
    _TCEFU_MUL32X8("MUL_ACCURATE", inA, inB, out);
}
else{
    _TCEFU_MUL32X8("MUL_AX", inA, inB, out);
}
\end{lstlisting}


\par As shown in Listing \ref{list:intrinsic}, the high-level compilation flow supports intrinsic driven operations, letting the user map a specific operation to a specific set of PEs. In this dummy example we can see that if the threshold value is true, the intrinsic {\fontfamily{qcr}\selectfont MUL\_ACCURATE}, is called, mapping the operation $out=in_A\cdot in_B$ to the accurate multiplication vector unit, and if the condition is false,, the same is applied for the approximate multiplication unit. The units in the provided example are 8 units supporting 32-bit operands as input ($32\times8$). In general cases, intrinsically driven mapping introduces a scheduling overhead, which can be minimized with the careful placement of the FUs on the CGRA grid and by removing any data dependencies from the high-level application.

\subsection{Voltage Island Formation}

\par The proposed CGRA is inherently heterogeneous, comprising of multiple tiles with varying characteristics in terms of area, power, and delay. For example, as shown in Table \ref{tab:drum}, the accurate multiplication blocks exhibit significantly higher delay (nearly twice as much) compared to their approximate counterparts. 
We leverage this observation to further enhance its performance and energy efficiency by reducing the operating voltage of CGRA tiles that exhibit greater timing slack relative to others. Operating these tiles at a lower voltage increases their delay due to slower data propagation, aligning their timing more closely with that of the critical tiles. Meanwhile, tiles with lower slack remain at the nominal voltage, preserving their performance. We choose to employ this static voltage scaling methodology, as opposed to more dynamic approaches like DVFS \cite{iced}, since profiling the DNN workloads on the finalized (post-pruning) CGRA architecture reveals extremely low variance in tile utilization, which does not compensate the extra area and switching cost for adopting DVFS over static voltage domains. This strategy not only reduces the total power consumption but also ensures that no timing violations occur, as voltage reduction is applied only to tiles with sufficient slack. Furthermore, the throughput of the system does not degrade, as the clock frequency is ultimately bounded by blocks that initially had the least slack (in our case, these are the $32\times32$ multipliers, which play a crucial role in address space computations). 

\par For our designs, we implement two voltage domains, i.e, one voltage domain at 0.6V incorporating the approximate multiplication tiles, the ALUs, the Register Files, and the switchboxes that are connected to these tiles, and a second one at 0.8V for the remaining tiles. While more fine-grain voltage islands are theoretically possible, (i.e., assigning a unique supply voltage to each tile), it can significantly increase design complexity, as it strains the physical design constraints, leading to challenges in the power delivery network design, routing congestion and area overhead due to the required placement of level shifters between different voltage regions\cite {level_shifter_overhead}.

\par The formation of the defined voltage regions was implemented with the use of Unified Power Format (UPF) descriptions to enable proper power-aware synthesis. Since the goal is to better balance the critical paths of the individual components (and tighten them whenever possible) by reducing the operating voltage of the DRUM multipliers and the ALUs, while maintaining the voltage of the accurate ones at the nominal value, we effectively decrease the timing slack deviation among the CGRA tiles from 300 psec to 104 psec.




\section{Mapping Framework}
\label{sec:approx_mapping}




\begin{figure*}[h!]
  \centering
  \includegraphics[width=\textwidth]{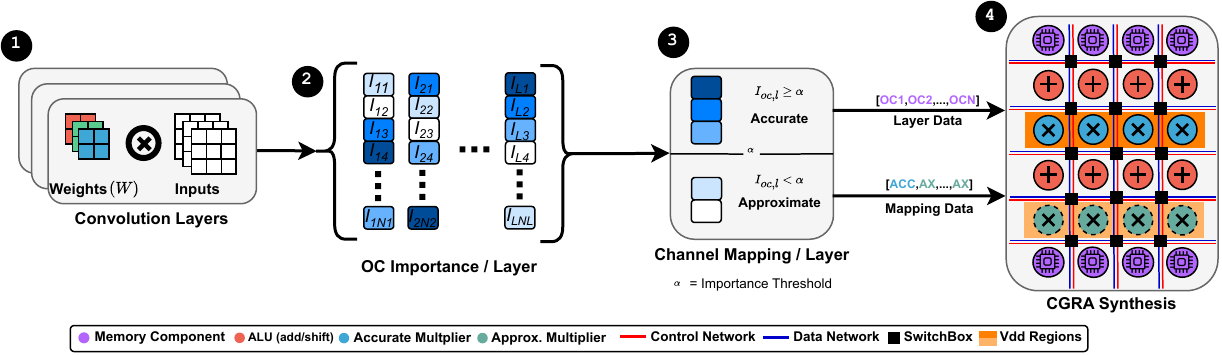}
  \caption{Overview of the proposed DNN mapping flow. For each conv layer, the process begins by extracting the importance score for each output channel. Based on these scores, each channel is assigned to either accurate or approximate units.}
  \label{fig:methodology}
  \vspace{-2ex}
\end{figure*}

This section describes the proposed design flow, detailing how DNN models are analyzed, optimized, and mapped onto the proposed approximate CGRA architecture. We formed a step-by-step methodology, shown in Figure \ref{fig:methodology}, starting from a high-level DNN model down to an optimized CGRA mapping that balances energy efficiency, computational accuracy, and hardware utilization.  

\subsection{Approximation-Aware DNN Model}


\par The process begins with a high-level DNN model, specifically implemented using the PyTorch framework and extended via the Xilinx Brevitas library \cite{brevitas} \circled{1}. Brevitas is a quantization-aware training library that facilitates the deployment of low-precision DNN models on hardware accelerators, making it well-suited for exploring approximate arithmetic effects. To incorporate approximate computing into our framework, we extend Brevitas to simulate the behavior of DRUM multipliers, ensuring an accurate representation of their impact on neural network computations. 
We extend the Brevitas framework by enabling the approximation of the features of a layer using dedicated approximate computing functional units. The behavior of the approximate functional units is described in the form of Look Up Tables (LUTs) that store the outcomes of all possible $N \times N$-bit approximate multiplications, 
e.g. enabling the evaluation of implications of DRUM$k$ over $N \times N$ multiplication,
allowing efficient value retrieval during inference.

\subsection{Importance Factors Calculation}
\par After extracting all the necessary information about our model's layers, we proceed with the calculation of the Importance Factors \circled{2}. The concept of importance estimation was originally introduced in the work of Molchanov et al. \cite{molchanov_importance_2019} as a fundamental principle for layer pruning in deep neural networks (DNNs). The authors calculate the importance of neurons (filters) by assessing the squared change in loss that results from their removal from the network. They approximate this calculation using Taylor expansions to ensure efficiency, especially in larger networks. The primary focus is its first-order approximation. More specifically, the importance score for a filter is defined as $I_m(W)=(g_mw_m)^2$, where $g_m$ represents the gradient available from backpropagation and $w_m$ is the corresponding weight of the filter. 

In our work, we adopt the same fundamental principle, but instead of performing kernel pruning, we extend the importance analysis to all convolutional kernels comprising each DNN filter. We propose calculating the Importance Factors for all kernels to determine which one should be mapped on the approximate components, rather than eliminating them from the network. This allows us to examine fine-grained allocation strategies to effectively assign different output channels within the same layer to either accurate or approximate multipliers. Implementing the above allocation strategy on top of a DNN dataflow that exploits output channel parallelism offers increased capabilities/opportunities for parallel execution without sacrificing the DNN's accuracy. For each layer, the Importance Factors can be calculated as follows:

\begin{equation}
    I_{oc,l} = MSE(Q_{out}(D,W),Q_{ax}(D,W,oc,l))
    \label{eq:importance}
\end{equation}

\noindent where $I_{oc,l}$ denotes the Importance of the output channel $oc$ in layer $l$, $Q_{out}$ refers to the output feature map that is produced when a set of input data $D$ is provided to a model with a weight set of $W$, and $Q_{ax}(D,W,oc,l)$ is the approximate output feature map generated, when the same input data $D$ and weights $W$ are given, but with approximate multiplications applied only on the $oc$ channel of layer $l$.

\subsection{Approximate vs. Accurate Kernel Mapping Strategy}
\par Once the Importance Factors ($I_{oc,l}$) have been calculated, the next step \circled{3} determines how each layer will be mapped on the CGRA. For our analysis, we consider a heterogeneous CGRA comprising both an approximate and an accurate computation region (see also Section \ref{subsec:compar_r_blocks}). This architectural configuration in conjunction with the output channel parallel DNN dataflow mentioned above, enables the concurrent execution of both accurate and approximate multiplications within a single clock cycle, thereby maximizing computational parallelism. 

The mapping process is structured into two distinct stages, as shown in Figure \ref{fig:methodology}. i) Sorting by Importance Factor: For each DNN layer, the output channels (OCs) are ranked in descending order based on their Importance Factors, prioritizing the most significant channels. ii) Approximate-Aware Mapping: Given a user-defined Quality-of-Service (QoS) constraint $\alpha$, the system assigns output channels to approximate multipliers starting from the least significant, progressively mapping additional channels until the QoS threshold is reached.
This structured mapping strategy ensures that computational resources are efficiently allocated, allowing approximate multipliers to be fully utilized while maintaining accuracy constraints dictated by the QoS parameter.

\par When all the layers are mapped efficiently onto the CGRA components, then the bitstream is produced, programming the accelerator and running the given kernel \circled{4}.
\section{Results \& Discussion}
\label{sec:evaluation}

\subsection{Experimental Setup}

\par In this section, we analyze the effectiveness of our work by performing an in-depth PPA evaluation over state-of-the-art CNNs. Since we primarily target Edge AI applications, for our analysis, we consider models commonly used in the computer vision model of MobileNetV2, which features multiple convolution layers with varying dimensions for sufficient benchmarking. To evaluate the impact of the approximate multipliers on the output features, we perform inference on the ImageNet dataset \cite{ImageNet}, as it is a challenging benchmark that effectively represents real-world scenarios.

\par To obtain measurements regarding area, power, and throughput, we implement our designs in Verilog HDL. For this experimental evaluation, we have built three different designs. The first is a relatively small scalar architecture that features four multipliers (one accurate, one approximate, and two for address generation and constant propagation) and four ALUs (Scalar). The remaining two are vector architectures featuring two vector lanes with widths of 4 (Vector-4) and 8 (Vector-8), respectively, meaning that 4 and 8 operations can be executed on the same clock cycle on each architecture. More specifically, the Vector-4 architecture includes 19 ALUs and multipliers in total, while the Vector-8 doubles this number.

\par The designs were synthesized using Synopsys Design Compiler, targeting a 22-nm technology node provided by GlobalFoundries, operating at 400 MHz under typical operating conditions (TT, 25\degree C). The Local and Instruction Memories were implemented using real SRAM Macros generated by the GlobalFoundries Memory Compiler. The synthesized netlists underwent functional validation by executing the aforementioned DNN benchmarks in Siemens QuestaSim. To estimate the overall power consumption, the switching activities of the netlists were extracted during the post-synthesis simulations, and subsequently, the resulting VCD files were converted to SAIF format for analysis using Synopsys PrimeTime.

\label{subsec:sota_comp}



\subsection{Approximation-Aware Analysis of MobileNetV2}

\begin{table}[h]
\centering
\small
\begin{tabular}{c|c|c|cc}
 &  &  & \multicolumn{2}{c}{OC map (\%)} \\
\multirow{-2}{*}{Quantile} & \multirow{-2}{*}{Perf (CC)} & \multirow{-2}{*}{RMSE} & Acc & Ax\\ \hline
1 & 52.7 M & 5.9 & 0 & 100.0 \\ \hline
0.875 & 49.7 M & 6.23 & 9.0 & 91.0 \\ \hline
0.75 & 46.1 M & 6.0 & 19.1 & 80.9 \\ \hline
0.5 & 40.7 M & 5.46 & 47.9 & 52.1 \\ \hline
0.25 & 46.1 M & 5.41 & 69.0 & 31.0 \\ \hline
0.125 & 49.6 M & 5.62 & 83.8 & 16.2 \\ \hline
0 & 52.7 M & 0 & 100.0 & 0
\end{tabular}
\caption{MobileNetV2 Performance-Error Analysis}
\label{tab:mbnv2}
\end{table}

\par We further explore how the user-defined QoS constraint $\alpha$ affects the output error; we evaluate different quantile thresholds of the Importance Factors for each output channel. This approach allows us to evaluate various partition strategies between accurate and approximate components, assessing their effect on computational accuracy and efficiency. We use quantile values of 0 (all accurate), 0.125, 0.25, 0.5 (median), 0.75, 0.875, and 1 (all approximate), with DRUM7 as the approximate multiplication component and the Vector-8 architecture. Table \ref{tab:mbnv2} shows the performance (in CCs) and output RMSE. As shown, the error propagation varies between different quantiles of importance due to different error propagation among layers, also providing energy trade-offs between mapping configurations. Finally, we observe that there is a linear relation between Acc-Ax mapping distribution and model performance, with the 0.5 quantile (nearly perfect split) representing the median value, showcasing the better performance.

\subsection{Comparison against Baseline R-Blocks CGRA}
\label{subsec:compar_r_blocks}

\begin{figure}[t]
    \centering
    \includegraphics[width=\columnwidth]{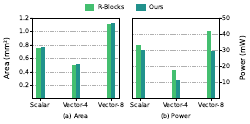}
    \caption{Area and power comparison of the examined architectures before (R-Blocks) and after DRUM Multiplier Integration and Voltage Scaling.}
    \label{fig:ppa}
    \vspace{-3ex}
\end{figure}

\par To assess the effectiveness of our methodology in terms of area and power efficiency, we perform a comparative study between scalar \& vectorized architectures and their iso-resource R-Blocks architecture. 
Figure \ref{fig:ppa} demonstrates the area measurements of the same scalar, Vector-4, and Vector-8 architectures built both in R-Blocks and our methodology. As expected, the scalar architecture exhibits the largest area overhead relative to the number of processing elements (PEs) included in each design. This is attributed to the increased number of Instruction Memories coupled alongside each PE. Scaling up the scalar architecture further amplifies this overhead, as the duplication of IMs becomes increasingly costly in terms of area. However, this limitation is mitigated in the vectorized architectures, where the vector units share common IMs, resulting in a  reduced silicon footprint. It is worth noting that the insertion of level shifters between different voltage domains has a very small impact on area, resulting in less than a 2\% increase. Furthermore, between the two vectorized architectures, we observe an almost linear increment in area, which aligns with the scaling of the computational resources.

\par A similar trend is observed in the measurements regarding the overall power consumption. Despite its lower computational capabilities, the scalar architecture demonstrates inferior power efficiency when compared to the vectorized designs, further highlighting the benefits of the vectorized implementations of the CGRAs. As shown in Figure \ref{fig:ppa}b, the integration of DRUM multipliers and the introduction of voltage islands around them significantly improves the power efficiency of the vectorized designs. Specifically, the total power consumption is reduced by 32.6\% in the Vector-4 architecture and by 29.3\% in the Vector-8 configuration. Interestingly enough, the smallest architecture shows minimal benefit from the inclusion of the DRUM multipliers and the proposed voltage scaling scheme (just a 6\% decrease in power), primarily due to the small number of approximate multipliers in the design (only one), which constrains the potential for larger power savings.

\subsection{Comparison with State-of-the-Art}

\par To finalize our evaluation, we compare our methodology with several SOtA CGRA architectures. We noted that X-CGRA \cite{akbari_x-cgra_2020}, GREEN \cite{green_2024} and CGRA4ML \cite{cgra4ml} do not consider the impact of the memory components in their measurements, which can significantly affect both area and power consumption. In contrast, our evaluation includes memory components, which, on average, account for 35\% of the total cell area and contribute approximately 30\% to the overall power consumption. Compared to other solutions, we can generate competitive CGRA architectures w.r.t. GOPS and GOPS/W metrics while offering design freedom at the operational level, allowing for the final CGRA to have a different number of heterogeneous PEs at different locations on the grid.
\section{Conclusions}
\label{sec:conclusion}
\par In this work, we present an end-to-end synthesis and mapping framework for approximate CGRA Edge AI processors. By employing a per-output channel mapping strategy that selectively assigns computations to approximate multipliers, optimizing energy efficiency while maintaining accuracy 
We analyzed our proposed framework on the widely used NN model for edge devices, MobileNetV2, showcasing that our methodology can generate architectures that can deliver up to 440 GOPS/W with minimal output error, outperforming several state-of-the-art CGRA architectures and frameworks both in terms of throughput and energy efficiency.

{
\bibliographystyle{IEEEtran}
\bibliography{references}

\begin{thebibliography}{10}
\providecommand{\url}[1]{#1}
\csname url@samestyle\endcsname
\providecommand{\newblock}{\relax}
\providecommand{\bibinfo}[2]{#2}
\providecommand{\BIBentrySTDinterwordspacing}{\spaceskip=0pt\relax}
\providecommand{\BIBentryALTinterwordstretchfactor}{4}
\providecommand{\BIBentryALTinterwordspacing}{\spaceskip=\fontdimen2\font plus
\BIBentryALTinterwordstretchfactor\fontdimen3\font minus \fontdimen4\font\relax}
\providecommand{\BIBforeignlanguage}[2]{{%
\expandafter\ifx\csname l@#1\endcsname\relax
\typeout{** WARNING: IEEEtran.bst: No hyphenation pattern has been}%
\typeout{** loaded for the language `#1'. Using the pattern for}%
\typeout{** the default language instead.}%
\else
\language=\csname l@#1\endcsname
\fi
#2}}
\providecommand{\BIBdecl}{\relax}
\BIBdecl

\bibitem{aisurvey}
\BIBentryALTinterwordspacing
X.~Wang, Z.~Tang, J.~Guo, T.~Meng, C.~Wang, T.~Wang, and W.~Jia, ``Empowering edge intelligence: A comprehensive survey on on-device ai models,'' \emph{ACM Comput. Surv.}, Mar. 2025, just Accepted. [Online]. Available: \url{https://doi.org/10.1145/3724420}
\BIBentrySTDinterwordspacing

\bibitem{edgetpu}
\BIBentryALTinterwordspacing
K.~Seshadri, B.~Akin, J.~Laudon, R.~Narayanaswami, and A.~Yazdanbakhsh, ``An evaluation of edge tpu accelerators for convolutional neural networks,'' 2022. [Online]. Available: \url{https://arxiv.org/abs/2102.10423}
\BIBentrySTDinterwordspacing

\bibitem{gemmini}
\BIBentryALTinterwordspacing
H.~Genc, S.~Kim, A.~Amid, A.~Haj-Ali, V.~Iyer, P.~Prakash, J.~Zhao, D.~Grubb, H.~Liew, H.~Mao, A.~Ou, C.~Schmidt, S.~Steffl, J.~Wright, I.~Stoica, J.~Ragan-Kelley, K.~Asanovic, B.~Nikolic, and Y.~S. Shao, ``Gemmini: Enabling systematic deep-learning architecture evaluation via full-stack integration,'' in \emph{2021 58th ACM/IEEE Design Automation Conference (DAC)}.\hskip 1em plus 0.5em minus 0.4em\relax IEEE Press, 2021, p. 769–774. [Online]. Available: \url{https://doi.org/10.1109/DAC18074.2021.9586216}
\BIBentrySTDinterwordspacing

\bibitem{akbari_x-cgra_2020}
\BIBentryALTinterwordspacing
O.~Akbari, M.~Kamal, A.~Afzali-Kusha, M.~Pedram, and M.~Shafique, ``X-{CGRA}: {An} {Energy}-{Efficient} {Approximate} {Coarse}-{Grained} {Reconfigurable} {Architecture},'' \emph{IEEE Transactions on Computer-Aided Design of Integrated Circuits and Systems}, vol.~39, no.~10, pp. 2558--2571, Oct. 2020. [Online]. Available: \url{https://ieeexplore.ieee.org/document/8815818/}
\BIBentrySTDinterwordspacing

\bibitem{green_2024}
\BIBentryALTinterwordspacing
Z.~Ebrahimi and A.~Kumar, ``{GREEN}: {An} {Approximate} {SIMD}/{MIMD} {CGRA} for {Energy}-{Efficient} {Processing} at the {Edge},'' \emph{IEEE Transactions on Computer-Aided Design of Integrated Circuits and Systems}, vol.~43, no.~10, pp. 2874--2887, Oct. 2024. [Online]. Available: \url{https://ieeexplore.ieee.org/document/10488043/}
\BIBentrySTDinterwordspacing

\bibitem{approxonthefly}
\BIBentryALTinterwordspacing
M.~Brandalero, L.~Carro, A.~C.~S. Beck, and M.~Shafique, ``Approximate on-the-fly coarse-grained reconfigurable acceleration for general-purpose applications,'' in \emph{Proceedings of the 55th Annual Design Automation Conference}, ser. DAC '18.\hskip 1em plus 0.5em minus 0.4em\relax New York, NY, USA: Association for Computing Machinery, 2018. [Online]. Available: \url{https://doi.org/10.1145/3195970.3195993}
\BIBentrySTDinterwordspacing

\bibitem{cgrame}
S.~A. Chin, N.~Sakamoto, A.~Rui, J.~Zhao, J.~H. Kim, Y.~Hara-Azumi, and J.~Anderson, ``Cgra-me: A unified framework for cgra modelling and exploration,'' in \emph{2017 IEEE 28th International Conference on Application-specific Systems, Architectures and Processors (ASAP)}, 2017, pp. 184--189.

\bibitem{revamp}
\BIBentryALTinterwordspacing
T.~K. Bandara, D.~Wijerathne, T.~Mitra, and L.-S. Peh, ``Revamp: a systematic framework for heterogeneous cgra realization,'' in \emph{Proceedings of the 27th ACM International Conference on Architectural Support for Programming Languages and Operating Systems}, ser. ASPLOS '22.\hskip 1em plus 0.5em minus 0.4em\relax New York, NY, USA: Association for Computing Machinery, 2022, p. 918–932. [Online]. Available: \url{https://doi.org/10.1145/3503222.3507772}
\BIBentrySTDinterwordspacing

\bibitem{cgra4ml}
\BIBentryALTinterwordspacing
G.~Abarajithan, Z.~Ma, Z.~Li, S.~Koparkar, R.~Munasinghe, F.~Restuccia, and R.~Kastner, ``Cgra4ml: A framework to implement modern neural networks for scientific edge computing,'' 2024. [Online]. Available: \url{https://arxiv.org/abs/2408.15561}
\BIBentrySTDinterwordspacing

\bibitem{mlcgra}
Y.~Luo, C.~Tan, N.~B. Agostini, A.~Li, A.~Tumeo, N.~Dave, and T.~Geng, ``Ml-cgra: An integrated compilation framework to enable efficient machine learning acceleration on cgras,'' in \emph{2023 60th ACM/IEEE Design Automation Conference (DAC)}, 2023, pp. 1--6.

\bibitem{iced}
C.~Tan, M.~Jiang, D.~Patil, Y.~Ou, Z.~Li, L.~Ju, T.~Mitra, H.~Park, A.~Tumeo, and J.~Zhang, ``Iced: An integrated cgra framework enabling dvfs-aware acceleration,'' in \emph{2024 57th IEEE/ACM International Symposium on Microarchitecture (MICRO)}, 2024, pp. 1338--1352.

\bibitem{de_bruin_r-blocks_2024}
\BIBentryALTinterwordspacing
B.~De~Bruin, K.~Vadivel, M.~Wijtvliet, P.~Jääskeläinen, and H.~Corporaal, ``\BIBforeignlanguage{en}{R-{Blocks}: an {Energy}-{Efficient}, {Flexible}, and {Programmable} {CGRA}},'' \emph{\BIBforeignlanguage{en}{ACM Transactions on Reconfigurable Technology and Systems}}, vol.~17, no.~2, pp. 1--34, Jun. 2024. [Online]. Available: \url{https://dl.acm.org/doi/10.1145/3656642}
\BIBentrySTDinterwordspacing

\bibitem{7818353}
M.~Wijtvliet, L.~Waeijen, and H.~Corporaal, ``Coarse grained reconfigurable architectures in the past 25 years: Overview and classification,'' in \emph{2016 International Conference on Embedded Computer Systems: Architectures, Modeling and Simulation (SAMOS)}, 2016, pp. 235--244.

\bibitem{opencgra}
C.~Tan, C.~Xie, A.~Li, K.~J. Barker, and A.~Tumeo, ``Opencgra: An open-source unified framework for modeling, testing, and evaluating cgras,'' in \emph{2020 IEEE 38th International Conference on Computer Design (ICCD)}, 2020, pp. 381--388.

\bibitem{hashemi_drum_2015}
\BIBentryALTinterwordspacing
S.~Hashemi, R.~I. Bahar, and S.~Reda, ``{DRUM}: {A} {Dynamic} {Range} {Unbiased} {Multiplier} for approximate applications,'' in \emph{2015 {IEEE}/{ACM} {International} {Conference} on {Computer}-{Aided} {Design} ({ICCAD})}.\hskip 1em plus 0.5em minus 0.4em\relax Austin, TX, USA: IEEE, Nov. 2015, pp. 418--425. [Online]. Available: \url{http://ieeexplore.ieee.org/document/7372600/}
\BIBentrySTDinterwordspacing

\bibitem{hycube}
M.~Karunaratne, A.~K. Mohite, T.~Mitra, and L.-S. Peh, ``Hycube: A cgra with reconfigurable single-cycle multi-hop interconnect,'' in \emph{2017 54th ACM/EDAC/IEEE Design Automation Conference (DAC)}, 2017, pp. 1--6.

\bibitem{mlir}
\BIBentryALTinterwordspacing
C.~Lattner, M.~Amini, U.~Bondhugula, A.~Cohen, A.~Davis, J.~Pienaar, R.~Riddle, T.~Shpeisman, N.~Vasilache, and O.~Zinenko, ``Mlir: A compiler infrastructure for the end of moore's law,'' 2020. [Online]. Available: \url{https://arxiv.org/abs/2002.11054}
\BIBentrySTDinterwordspacing

\bibitem{wilton}
S.~A. Razavi, M.~S. Zamani, and K.~Bazargan, ``A tileable switch module architecture for homogeneous 3d fpgas,'' in \emph{2009 IEEE International Conference on 3D System Integration}, 2009, pp. 1--4.

\bibitem{jaaskelainen2017hw}
P.~J{\"a}{\"a}skel{\"a}inen, T.~Viitanen, J.~Takala, and H.~Berg, ``Hw/sw co-design toolset for customization of exposed datapath processors,'' \emph{Computing platforms for software-defined radio}, pp. 147--164, 2017.

\bibitem{tta}
H.~Corporaal and R.~Lamberts, ``Tta processor synthesis,'' in \emph{First Annual Conf. of ASCI}.\hskip 1em plus 0.5em minus 0.4em\relax Citeseer, 1995, pp. 18--27.

\bibitem{microAX18}
V.~Leon, G.~Zervakis, S.~Xydis, D.~Soudris, and K.~Pekmestzi, ``Walking through the energy-error pareto frontier of approximate multipliers,'' \emph{IEEE Micro}, vol.~38, no.~4, pp. 40--49, 2018.

\bibitem{dacAX19}
\BIBentryALTinterwordspacing
V.~Leon, K.~Asimakopoulos, S.~Xydis, D.~Soudris, and K.~Pekmestzi, ``Cooperative arithmetic-aware approximation techniques for energy-efficient multipliers,'' in \emph{Proceedings of the 56th Annual Design Automation Conference 2019}, ser. DAC '19.\hskip 1em plus 0.5em minus 0.4em\relax New York, NY, USA: Association for Computing Machinery, 2019. [Online]. Available: \url{https://doi.org/10.1145/3316781.3317793}
\BIBentrySTDinterwordspacing

\bibitem{Jiang2020}
H.~Jiang, F.~J.~H. Santiago, H.~Mo, L.~Liu, and J.~Han, ``Approximate arithmetic circuits: A survey, characterization, and recent applications,'' \emph{Proceedings of the IEEE}, vol. 108, no.~12, pp. 2108--2135, 2020.

\bibitem{TECS24Axsurvey}
\BIBentryALTinterwordspacing
Y.~Wu, C.~Chen, W.~Xiao, X.~Wang, C.~Wen, J.~Han, X.~Yin, W.~Qian, and C.~Zhuo, ``A survey on approximate multiplier designs for energy efficiency: From algorithms to circuits,'' \emph{ACM Trans. Des. Autom. Electron. Syst.}, vol.~29, no.~1, Jan. 2024. [Online]. Available: \url{https://doi.org/10.1145/3610291}
\BIBentrySTDinterwordspacing

\bibitem{level_shifter_overhead}
M.~R. Kakoee and L.~Benini, ``Fine-grained power and body-bias control for near-threshold deep sub-micron cmos circuits,'' \emph{IEEE Journal on Emerging and Selected Topics in Circuits and Systems}, vol.~1, no.~2, pp. 131--140, 2011.

\bibitem{brevitas}
AMD-Xilinx, ``Brevitas,'' \url{https://github.com/Xilinx/brevitas}.

\bibitem{molchanov_importance_2019}
\BIBentryALTinterwordspacing
P.~Molchanov, A.~Mallya, S.~Tyree, I.~Frosio, and J.~Kautz, ``Importance {Estimation} for {Neural} {Network} {Pruning},'' Jun. 2019, arXiv:1906.10771 [cs]. [Online]. Available: \url{http://arxiv.org/abs/1906.10771}
\BIBentrySTDinterwordspacing

\bibitem{ImageNet}
J.~Deng, W.~Dong, R.~Socher, L.-J. Li, K.~Li, and L.~Fei-Fei, ``Imagenet: A large-scale hierarchical image database,'' in \emph{2009 IEEE Conference on Computer Vision and Pattern Recognition}, 2009, pp. 248--255.

\end{thebibliography}
}

\end{document}